\documentclass[twocolumn]{aastex61}

\newcommand\aastex{AAS\TeX}

\received{July 1, 2016}
\revised{September 27, 2016}
\accepted{\today}
\submitjournal{ApJS}

\shorttitle{\aastex\ sample article}
\shortauthors{Ku\'zmicz \& Jamrozy}

\begin{document}

\title{Giant radio quasars: sample and basic properties}

\correspondingauthor{Agnieszka Ku\'zmicz}
\email{cygnus@oa.uj.edu.pl}

\author[0000-0002-3097-5605]{Agnieszka Ku\'zmicz}
\affil{Astronomical Observatory, Jagiellonian University, ul. Orla 171, 30-244 Krakow, Poland}
\affil{Queen Jadwiga Astronomical Observatory in Rzepiennik Biskupi, 33-163 Rzepiennik Strzy\.zewski, Poland}

\author{Marek Jamrozy}
\affiliation{Astronomical Observatory, Jagiellonian University, ul. Orla 171, 30-244 Krakow, Poland}



\begin{abstract}

We present the largest sample of giant radio quasars (GRQs), which are defined as having a projected linear size greater than 0.7 Mpc. The sample consists of 272 GRQs, of which 174 are new objects discovered through cross-matching the NRAO VLA Sky Survey (NVSS) and the Sloan Digital Sky Survey 14$^{\rm th}$ Data Release Quasar Catalogue (DR14Q) and confirmed using Faint Images of the Radio Sky at Twenty-Centimeters (FIRST) radio maps. In our analysis we compare the GRQs with 367 smaller, lobe-dominated radio quasars found using our search method, as well as with quasars from the SDSS DR14 Quasar Catalogue, investigating the parameters characterizing their radio emission (i.e. total and core radio luminosity, radio core prominence), optical properties (black hole masses, accretion rates, distribution in Eigenvector 1 plane) and infrared colours. For the GRQs and smaller radio quasars we find a strong correlation between [OIII] luminosity and radio luminosity at 1.4 GHz, indicating a strong connection between radio emission and conditions in the narrow-line region. We spot no significant differences between GRQs and smaller radio quasars, however we show that most extended radio quasars belong to a quasar population of evolved AGNs with large black hole masses and low accretion rates.We also show that GRQs have bluer W2-W3 colours compared to SDSS quasars with FIRST detections, indicating differences in the structure of the dusty torus.

\end{abstract}


\section{Introduction} \label{sec:intro}

Giant radio sources (GRSs) are objects with extremely large projected linear sizes ($>$ 0.7 Mpc; assuming H$_0$=71 km $\rm s^{-1} Mpc^{-1}$, $\Omega_M$=0.27, $\Omega_{\Lambda}$=0.73; \citealt{spergel2003}). It is believed that such large radio size structures are relatively rare. Only $\sim$6\% of radio sources from the 3CR complete sample do exceed this size (e.g. \citealt{ishwara99}). The reasons why some radio sources have grown so large are not fully understood, however detailed multi-wavelength studies have significantly increased our knowledge about the nature of GRSs (e.g. \citealt{jamrozy2008}, \citealt{machalski2009}, \citealt{konar2008}, \citealt{kuligowska2009}, \citealt{subrahmanyan2008}).

The crucial point in research of the GRS' origin is to study large and homogeneous samples of such objects. Owing to the efforts of many scientists, a~lot of new GRSs were found during the last several years. \cite{kuzmicz2018} catalogued all the GRSs found in the literature up to 2018. The sample includes 349 GRSs, of which 280 are hosted by galaxies (giant radio galaxies; GRGs) and 69 by quasars (giant radio quasars; GRQs). The second-largest sample of GRSs was compiled by \cite{dabhade2020}. Based on the low-frequency LOFAR Two-metre Sky Survey first data release \citep[LoTSS;][]{shimwell2017, shimwell2019}, the authors collected 239 GRSs (199 GRGs and 40 GRQs). This release covers a region of only 424 deg$^2$, but the LOFAR survey is very sensitive to low surface brightness features and has a high angular resolution, which makes it a valuable tool in identifying extended radio galaxies. A smaller sample of GRSs was also compiled by \cite{koziel2020}, as part of the ROGUE project within which the authors catalogued 33 GRGs.
Recently, a new large catalogue is being prepared by \cite{dabhade2020b} under the SAGAN project, where authors collected 162 GRSs (139 GRGs and 23 GRQs) using NVSS (\citealt{condon1998}), FIRST (\citealt{becker1995}), and the TIFR GMRT Sky Survey (TGSS; \citealt{intema2017}). In total, we know at least about 770 GRSs, of which only 109 are considered as GRQs. In the 3CRR complete sample of radio sources with flux density limit 10 Jy \citep{laing1983}, 75\% of radio sources are radio galaxies, and 25\% are radio quasars, according to the NASA Extragalactic Database (NED, \url{ned.ipac.caltech.edu}). The smaller fraction of GRQs in the entire GRS population indicates that selection effects play a significant role in their identification.

The connection between the production of powerful jets and the conditions within the innermost regions of active galactic nuclei (AGN) has not been fully explored. Various studies have attempted to understand the physical processes underlying the optical and radio emission in quasars (e.g. \citealt{jackson1991}, \citealt{willott1999}, \citealt{miller1999}, \citealt{sulentic2002}, \citealt{kimball2011}, \citealt{jackson2013}, \citealt{olmo2020}, \citealt{gaur2019}). A lot of research has concentrated on testing unified schemes where radio-loud quasars are supposedly galaxies with a central supermasive black hole (BH) surrounded by an accretion disk, a dusty torus and clouds of gas. They can generate powerful radio jets directed along the rotation axis of the BH. The anisotropic emission due to torus obscuration as well as relativistic boosting of radio jet emission leads to the orientation effects visible in optical and radio bands. Certain spectral parameters were found to correlate with the radio source's orientation. For example, \cite{kharb2004} found a correlation between nuclear optical luminosity and radio core prominence, which can be used as an orientation indicator. Also, the equivalent widths of broad emission lines are found to be orientation-dependent e.g. \cite{baker1997} and \cite{kimball2011}, although the authors obtained contradictory results claiming respectively anticorrelation and correlation between those two parameters. However, the basic AGN model predicts a stronger continuum emission in sources viewed closer to the radio-jet axis, which leads to smaller values of equivalent widths. \\
The connection between the extended radio luminosity and luminosities of narrow emission lines found by different authors (e.g. \citealt{baum1989}, \citealt{rawlings1989}, \citealt{tadhunter1998}, \citealt{gaur2019}) indicates that the source responsible for narrow-line emission is actually also the source of radio emission \citep{rawlings1991}, in contrast to other models in which radio and narrow-line luminosities are mainly driven by the environment \citep{dunlop1993}. 
The narrow-line and radio emission are most likely related to the accretion rate and/or the BH mass, which are also drivers of AGN evolution.\\
The GRSs that are supposed to be sources in an advanced evolutionary stage, were studied in terms of their optical properties by \cite{kuzmicz2012}. The authors analysed BH masses, accretion rates, and radio properties for a sample of 45 GRQs.  
As a result, they discovered that GRQs are very similar to smaller radio quasars and the determined parameters are typical for powerful quasars. It has to be noted that the analysed sample was relatively small as compared to the number of GRQs known to date. Therefore, we decided to re-examine optical properties of GRQs, focusing on various aspects which had not been explored yet.

The aim of this work is to complement the existing samples of GRSs with new objects which are hosted by quasars. While the number of known GRGs ($\sim$650) is relatively high, the GRQs constitute only $\sim$14\% of all known GRSs. It is much less than in the 3CRR sample where the QSOs constitute 25\% of radio sources. It has to be emphasized that our sample enlarges the number of known GRQs nearly threefold, which shows our method to be an efficient way of finding GRSs. The second part of this study concentrates on some fundamental properties observed in AGNs, i.e. the quasar main sequence, infrared colour diagram, and radio core prominence. We compare properties of GRQs with smaller-sized extended radio quasars (SRQs), as well as with the SDSS quasars that have matches with a FIRST radio source according to the SDSS data release 14 Quasar catalogue (DR14Q; \citealt{paris2018}) to look for differences between the GRQs and other quasars.

\section{Sample selection}

For our analysis, we collected GRQs known to date, along with 174 new objects, which makes our sample the largest one containing such a rare class of radio sources. 69 GRQs out of all previously known GRQs were taken from the literature compilation of GRSs by \cite{kuzmicz2018}. A further 15 GRQs were taken from \cite{dabhade2020}, where authors identified them on low frequency radio images of the LoTSS survey, and another 14 GRQs were taken from \cite{dabhade2020b}. 
The new GRQs studied in this paper were found using the currently available data from the NVSS, FIRST and DR14Q catalogues. These catalogues were already used in a systematic search of GRSs (e.g. \cite{proctor2016}, \cite{dabhade2017}, \cite{dabhade2020b}).
However, e.g. \cite{proctor2016} selected only radio sources with angular size larger than 4$^\prime$. In our study we do not apply any restriction on angular size and we found 59 new GRQs with angular sizes larger than 4$^\prime$. Therefore, in order to find extended radio sources, future search methods have to be improved.\\
The combined sample of new and known GRQs comprises 272 objects, which significantly enlarges the number of known GRQs.
The newly discovered 174 GRQs and their basic parameters are listed in Table 1. 

\subsection{GRQ search method}
\label{search}

\begin{figure}
\centering
    \includegraphics[width=0.99\columnwidth]{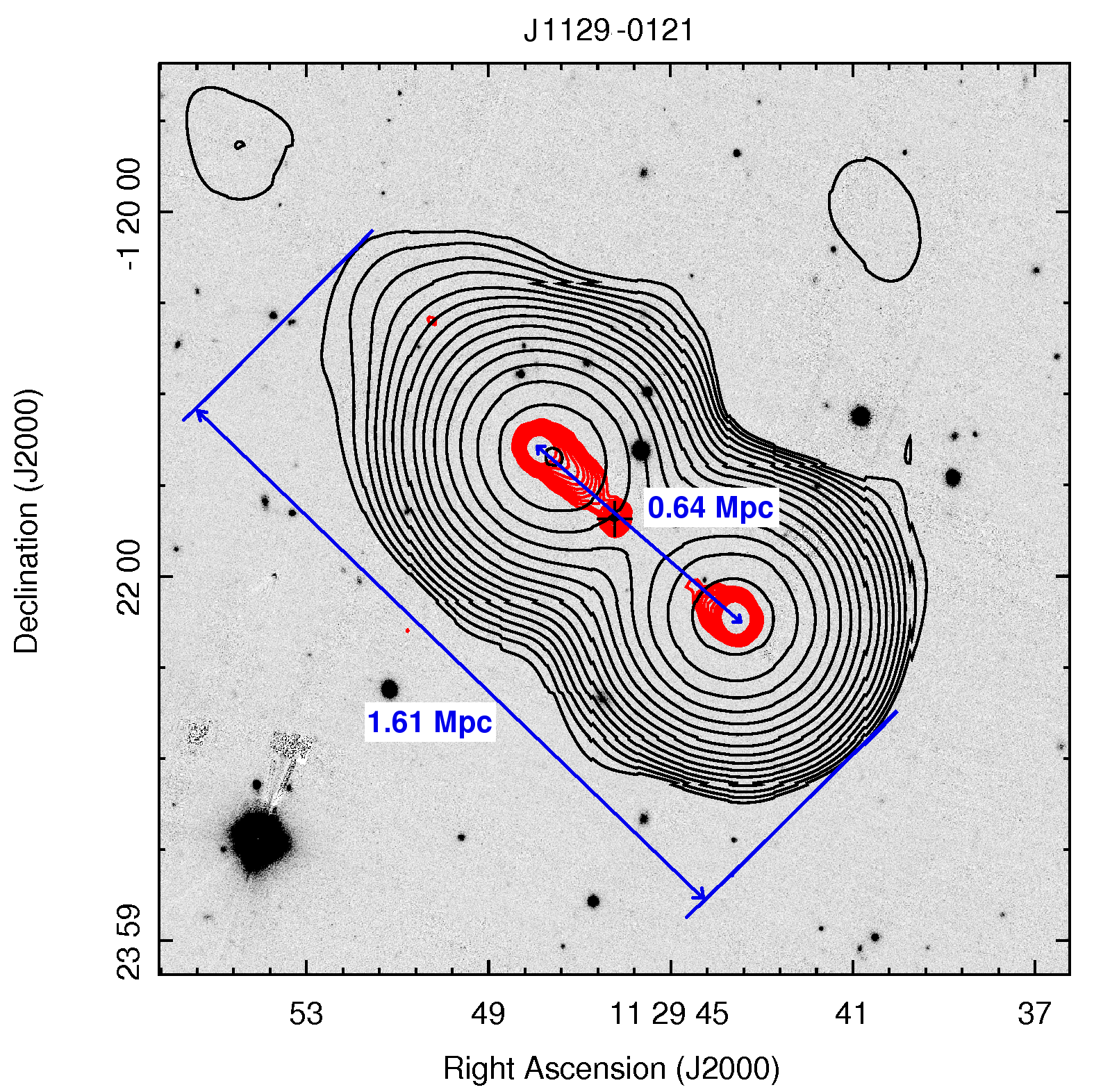}\\
\caption{The GRQ J1129-0121 located at z=0.726. The NVSS black contours and FIRST red contours are overlaid on an r-band Pan-STARRS optical image. The Figure shows the difference between radio source size measurement methods. In \cite{dabhade2020b} J1129-0121 is measured to 3$\sigma$ NVSS contour level giving the largest linear size D=1.61 Mpc. The measurement from FIRST hot-spot to hot-spot results the D=0.64 Mpc, disqualifying this quasar as a GRQ. In our study the J1129-0121 is classified as SRQ.}
\label{1129}
\end{figure}

In searching extended radio quasars we used the DR14Q quasar catalogue, including 526 356 spectroscopically identified quasars and quasar candidates. All of the catalogued quasars were cross-matched with the NVSS radio sources in the following way:
\begin{itemize}
\item In the first step, using the DS9 software\footnote{\url{http://ds9.si.edu}} we plotted the positions of all quasars from the DR14Q on the full NVSS atlas images of 4$^\circ$$\times$4$^\circ$ in size. Only $\sim$1000 NVSS images had DR14Q objects in them and typically each NVSS image had about 500 such quasars. Around each quasar position we drew a circle corresponding to the expected 0.7 Mpc angular size, considering the redshift of each quasar.
\item In the second step we visually inspected all the full NVSS atlas images containing DR14Q objects, looking for radio sources exceeding the size of the plotted circles. We considered only the sources where the quasar position was near the centre of radio emission i.e. in a centre of single apparent elongated emission, between two maxima of radio emission, or in a maximum of radio emission between two nearly symmetrically located maxima. Due to such a selection strategy we could have missed very asymmetric radio sources and sources with a one-sided radio lobe visible in NVSS maps. As a result we selected 1341 quasars with visible elongated NVSS radio emission exceeding 0.7 Mpc in size.
\item In the next step we manually verified all positive matches using FIRST and AllWISE infrared images \citep{cutri2013} and SDSS optical images to discern false findings. We visually inspected whether radio hot-spots coincide with optical or infrared sources. We also checked if the positions of optical quasar hosts coincide with the FIRST radio core emissions. Almost all the quasars from our sample have radio cores in the FIRST survey catalogue separated by less than 1$^{\prime\prime}$ from the optical quasar. We confirmed that 603 out of 1341 selected quasars are the hosts of extended radio sources. Based on the FIRST radio maps, we measured the projected linear size (distance between the opposite hot-spots) of GRQ candidates. As a result, a lot of radio quasars proved to be actually smaller than 0.7 Mpc, despite their projected linear size on the NVSS maps (as measured to the 3$\sigma$ contour level -- step two of the search method) exceeding 0.7 Mpc (Section \ref{smaller}). In some cases the difference between both these methods of radio source size measurement, from one hot-spot to opposite hot-spot on the FIRST maps and from 3$\sigma$ contour level of one lobe to 3$\sigma$ contour level of the opposite lobe on the NVSS maps, is very large. This can be particularly well seen in the case of high-redshift quasars that have small angular sizes (e.g. for redshift z=1 the NVSS beam size equal to 45$^{\prime\prime}$ corresponds to $\sim$360 kpc), for which NVSS 3$\sigma$ sizes are overestimated more than twofold. Therefore, in our study we use radio source size measurements from the hot-spot to hot-spot method on the FIRST maps, while NVSS 3$\sigma$ sizes where used only in the process of GRQs candidates selection. The difference between the two methods of radio source size measurement is illustrated in Figure \ref{1129}. For the radio sources with no radio core emission detected in the FIRST survey, we checked the host position using the Very Large Array Sky Survey (VLASS; \citealt{lacy2020}) all-sky radio survey at 3 GHz with high angular resolution ($\sim$2.$^{\prime\prime}$5). We also used VLASS maps to determine angular sizes of radio sources which are outside the FIRST survey footprint or have FIRST radio structure visible only on one side of the host quasar. They are marked with letter ``v'' in Tables 1 and 2. In the cases where there was no possibility to measure the radio source's size from hot-spot to hot-spot (because there were no FIRST or VLASS detections), we give an approximated value measured between NVSS maxima of lobe emission (marked as ``nn'' in Tables 1 and 2) or between the FIRST hot-spot and the NVSS maximum on the opposite side in the case of one-sided FIRST radio lobes (marked as ``fn'' in Tables 1 and 2). In our study we use only the radio source size measurements listed in Tables 1 and 2.
\end{itemize}

It is worth to highlight that using the method described above we ``re-discovered'' almost all the GRQs from \cite{kuzmicz2018} within the field of the NVSS/SDSS surveys. Moreover, we found 15 out of 40 GRQs which have been found earlier by \cite{dabhade2020} based on the low-frequency LoTSS radio maps.
A further 19 GRQs from \cite{dabhade2020} are too small to meet the GRS size criterion, based on measurements of their radio structures on FIRST maps, therefore we do not include them in the sample of known GRQs. Of the 23 new GRQs found by \cite{dabhade2020b}, 6 GRQs were identified also in our search method, and 11 occurred to have hot-spot to hot-spot FIRST sizes smaller than 0.7 Mpc. These 11 QSOs were also not included in the sample of known GRQs. It has to be noted here that in \cite{dabhade2020} the authors measured sizes of radio sources up to the 3$\sigma$ level on low resolution (20$^{\prime\prime}$) LOFAR maps, and in \cite{dabhade2020b} up to the 3$\sigma$ level on NVSS maps, therefore their results are overestimated. In our study, for the quasars taken from \cite{dabhade2020} and \cite{dabhade2020b} we use remeasured sizes from hot-spot to hot-spot on FIRST maps. 

\subsection{Subsample of smaller radio quasars}
\label{smaller}

The method described above allowed for identification of extended radio quasars with NVSS 3$\sigma$ sizes larger than 0.7 Mpc. As was mentioned in Section \ref{search}, some of the radio quasars selected in the second step of the search method proved to be smaller than 0.7 Mpc after remeasuring their sizes from hot-spot to hot-spot in the FIRST radio maps. 
In our study such quasars are classified as SRQs.

In the SRQ sample we collected 367 objects that we found using our search method. They are listed in Table 2. Their projected linear sizes are between 0.2 -- 0.7 Mpc, so they are not GRQs but represent the population of extended radio quasars which can be used in other studies. Together with GRQs they provide a sample covering a continuous size range of smaller and larger radio quasars. The number of quasars in SRQ sample (with 3$\sigma$ sizes larger than 0.7 Mpc and hot-spot to hot-spot sizes smaller than 0.7 Mpc) shows that in samples where the 3$\sigma$ method is used for size measurement, radio source sizes can be even triply overestimated.

\subsection{Characteristics of the sample}

\begin{figure}
\centering
    \includegraphics[width=0.99\columnwidth]{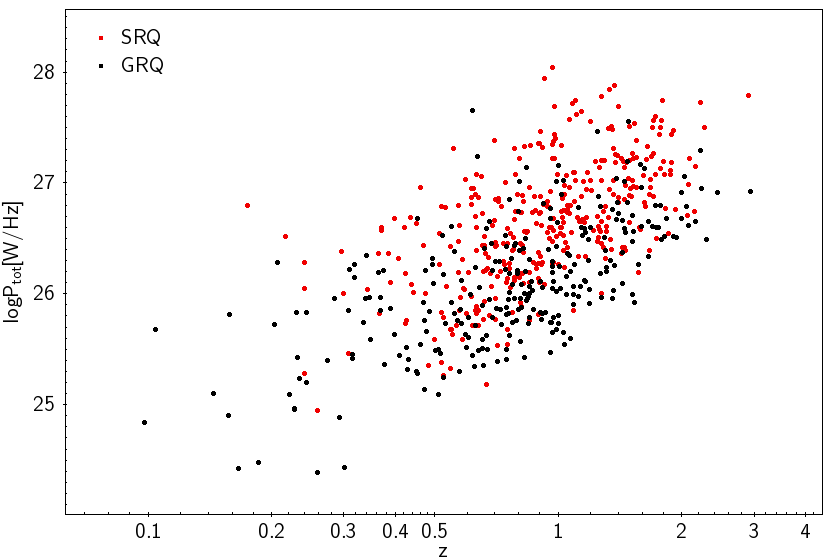}\\
\caption{The total radio luminosity at 1.4 GHz as a function of redshift for the GRQ (black dots) and SRQ (red dots) samples.}
\label{Pz}
\end{figure}

\begin{figure}
\centering
    \includegraphics[width=0.99\columnwidth]{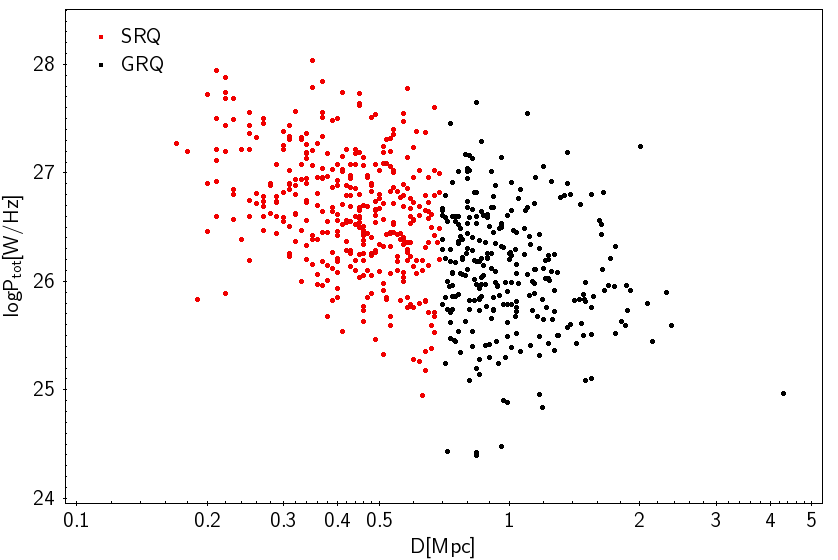}\\
\caption{Luminosity -- linear size diagram for GRQs and SRQs. The symbols are the same as in Figure \ref{Pz}. In the figure we use the linear size measured as the distance between hot-spots.}
\label{PD}
\end{figure}

The characteristic parameters of the GRQ (174 new findings and 98 previously known GRQs listed in \citealt{kuzmicz2018}, \citealt{dabhade2020} and \citealt{dabhade2020b}) and SRQ samples are presented in Figure \ref{Pz} and \ref{PD}, where we plot the redshift versus 1.4 GHz total radio luminosity (P$_{\rm tot}$) and total radio luminosity -- linear size diagram, respectively. 
The P$_{\rm tot}$ for newly identified GRQs and SRQs was determined by applying the formula given by \cite{brown2001}, where we used the 1.4 GHz flux-densities measured on the NVSS maps, adopting the spectral index value $\alpha=-0.6$ (after \citealt{wardle1997}). We use the convention of S$_{\nu}\sim \nu^{\alpha}$. In order to estimate the core radio luminosity (P$_{\rm core}$) analysed in the next sections, we measured core flux densities on the FIRST maps and used $\alpha=-0.3$ from \cite{zhang2003}. 


\begin{figure}
\centering
    \includegraphics[width=0.99\columnwidth]{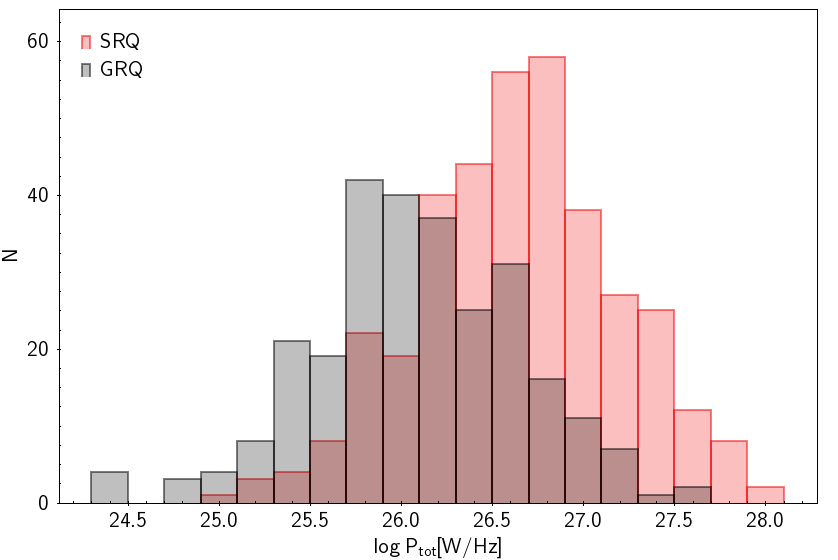}\\
    \includegraphics[width=0.99\columnwidth]{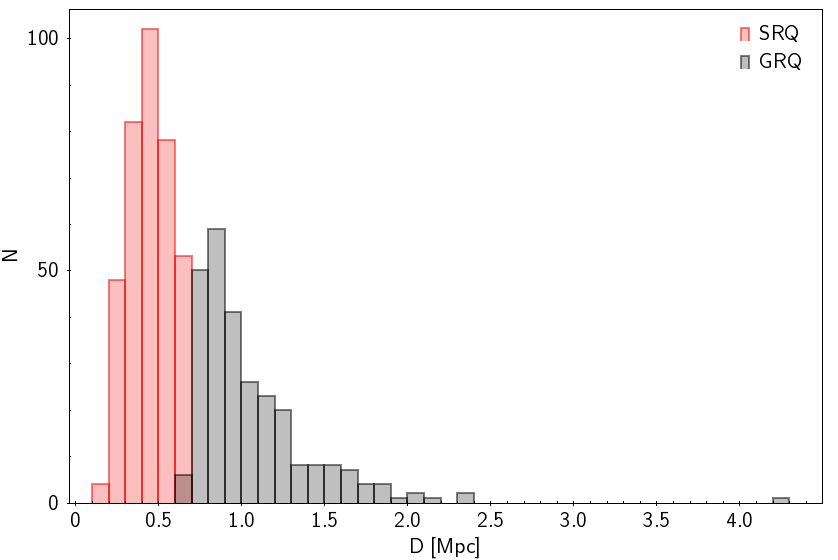}\\
\caption{Distribution of P$_{\rm tot}$ (top panel) and D (bottom panel) for GRQ and SRQ samples.}
\label{distr}
\end{figure}

\begin{figure}
\centering
    \includegraphics[width=0.99\columnwidth]{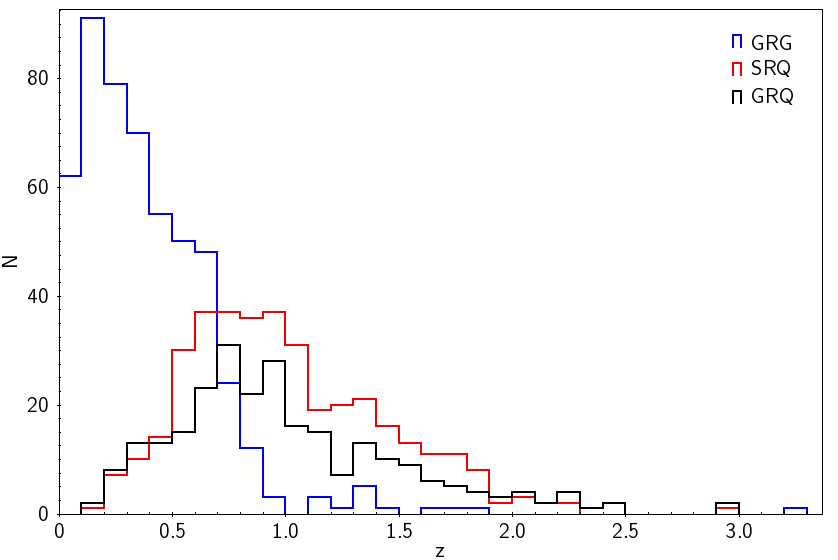}\\
\caption{Distribution of redshift for GRQs (black line), SRQs (red line) and GRGs (blue line).}
\label{distr_z}
\end{figure}

The quasars from our samples cover the redshift range of 0.1$<$z$<$3 and the median value of projected linear size (D) for GRQs is D=0.9 Mpc and for SRQs D=0.44 Mpc. The median value of 1.4 GHz total radio luminosity is $\log$P$_{\rm tot}[\rm W Hz^{-1}]$=26.1  for the GRQ sample and $\log$P$_{\rm tot}[\rm W Hz^{-1}]$=26.6 for the SRQ sample. The smaller median value of P$_{\rm tot}$ for GRQs is in agreement with existing radio source evolutionary models (e.g. \citealt{kaiser1997}), where the larger radio sources are older and thus have lower total radio luminosities. It can be clearly seen in Figure \ref{PD}, where we plotted P$_{\rm tot}$ against the projected linear size measured from hot-spot to hot-spot (column 6 in Table 1), that P$_{\rm tot}$ decreases as the projected linear size increases. The distributions of P$_{\rm tot}$ and D for both the samples are presented in Figure \ref{distr}. The largest GRQ J0931+3204 measures 4.29 Mpc (\citealt{coziol2017}).\\ 
In Figure \ref{distr_z}, we plot the distribution in redshift for our samples in comparison with the distribution for GRSs hosted by galaxies. It can be seen that the highest number of GRQs and SRQs is at the redshift z$\sim$0.8, while for GRGs the maximum of distribution is at z=0.2. The differences in redshift distributions for GRQs and GRGs are caused by selection effects. The galaxies at higher redshifts are hard to observe, while the highest number of SDSS DR14 quasars is observed at z$\sim$1.5 and z$\sim$2.2 with {only $\sim$50 objects below z=0.1.}

\begin{itemize}
\item High redshift (z$>$1) GRSs are very rare because the IGM density is higher at earlier cosmological epochs and also because the surface brightness of the radio structure strongly depends on redshift, which makes them hard to be detected and identified. Until now, 31 GRQs at z$>$1 of which only 6 have z$>$2 had been reported in literature \citep{kuzmicz2018, kuligowska2018, dabhade2020}. In the presented sample of GRQs, there are 70 new QSOs at z$>$1, of which 9 are located at z$\geq$2. This significantly increases the number of known high-redshift GRSs. The radio maps of the most distant newly discovered GRQs are presented in Figure \ref{z}, where we plotted NVSS and FIRST or VLASS contours overlaid onto r-band Pan-STARRS \citep{flewelling2020} optical image. The most distant GRQ is J1411+0156 located at z=2.95. As can be seen on the radio map of J1411+0156, there is no radio bridge connecting the radio core and the western radio lobe. Moreover, there is a very faint infrared object in the unWISE Catalog \citep{schlafly2019} which coincides with the western radio lobe. Therefore better sensitivity radio data are needed to full confirmation this radio quasar as a GRQ.

\begin{figure*}
\centering
    \includegraphics[width=0.68\columnwidth]{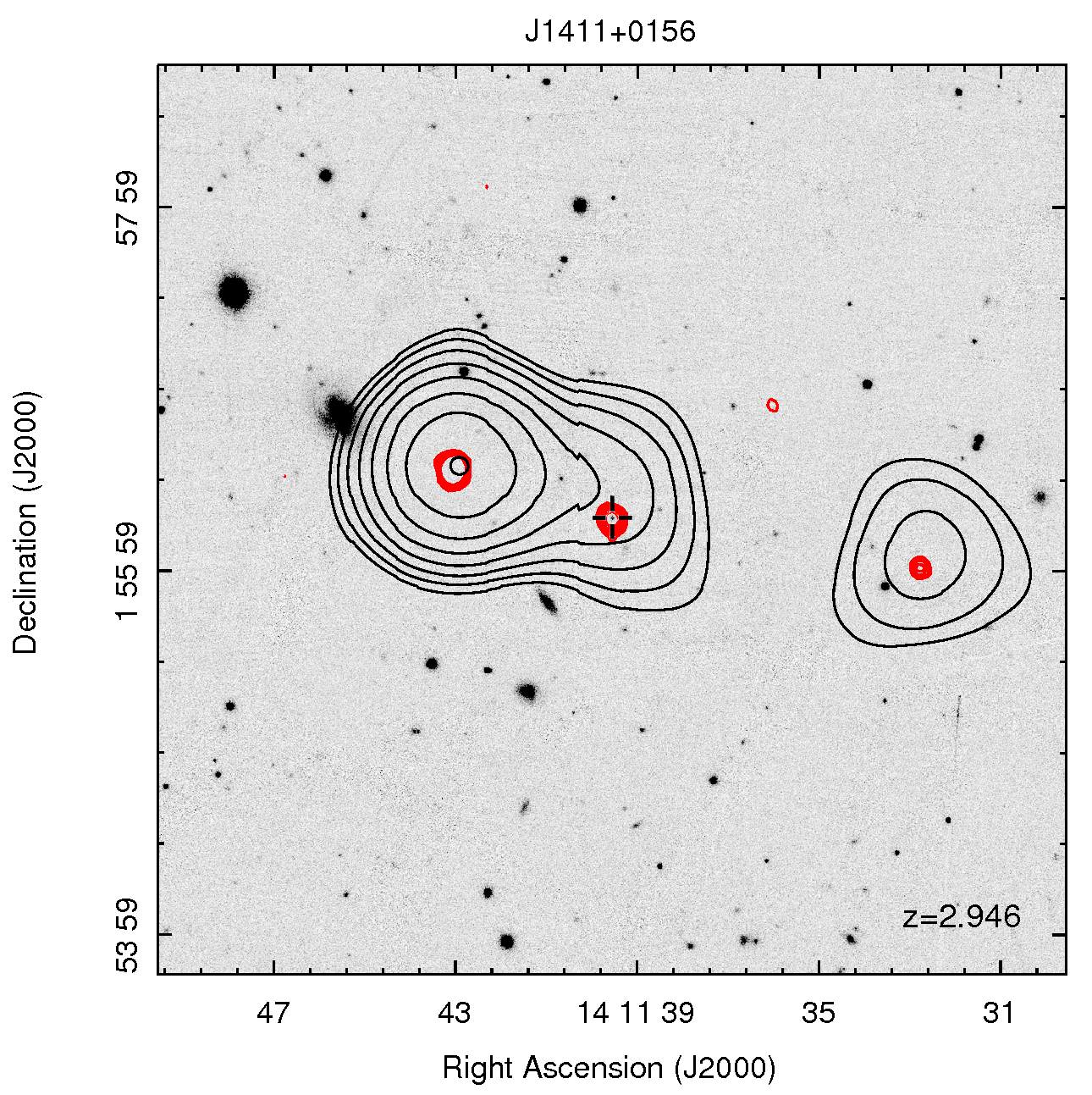}
    \includegraphics[width=0.68\columnwidth]{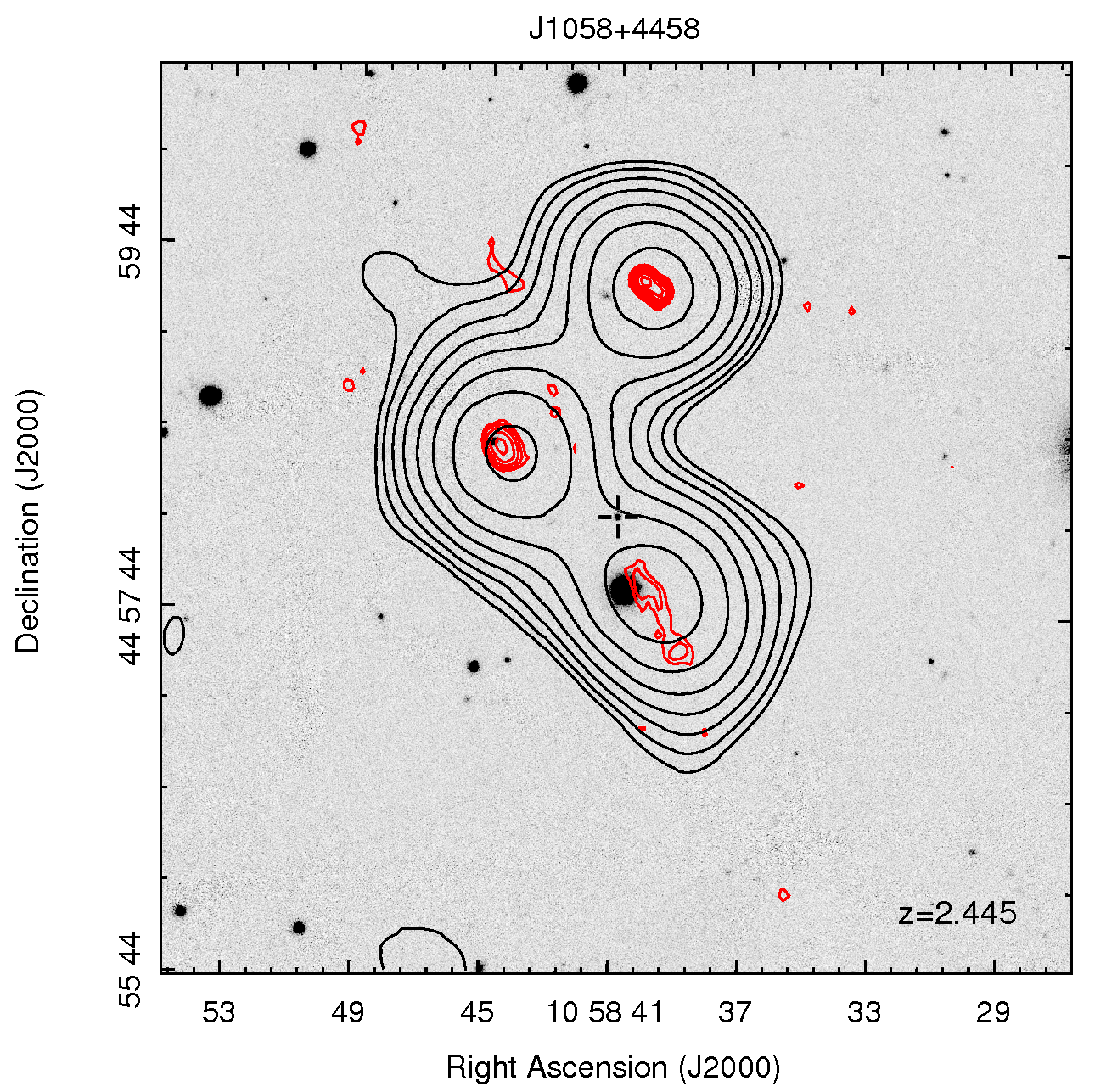}
    \includegraphics[width=0.68\columnwidth]{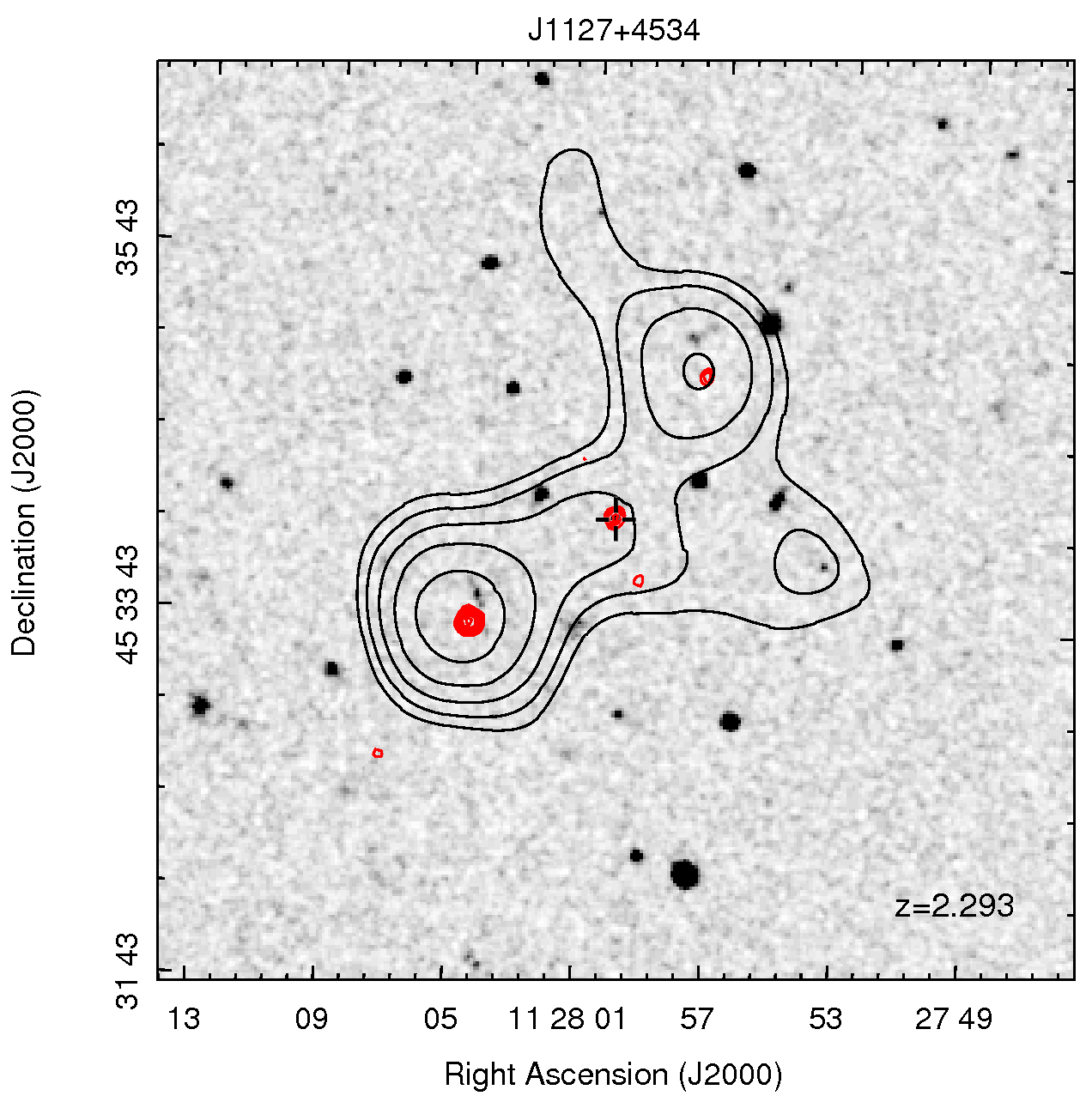}
    \includegraphics[width=0.68\columnwidth]{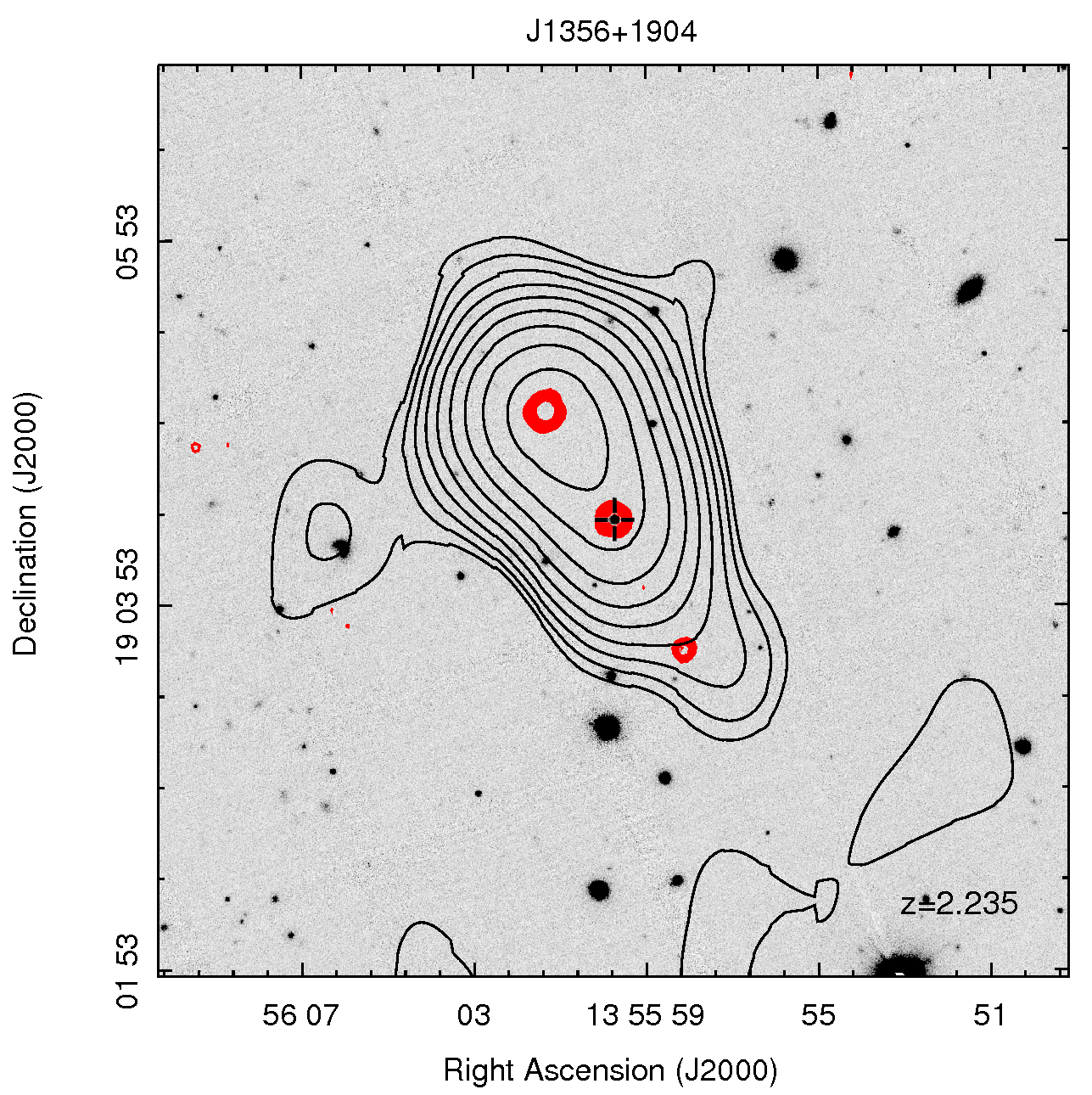}
    \includegraphics[width=0.68\columnwidth]{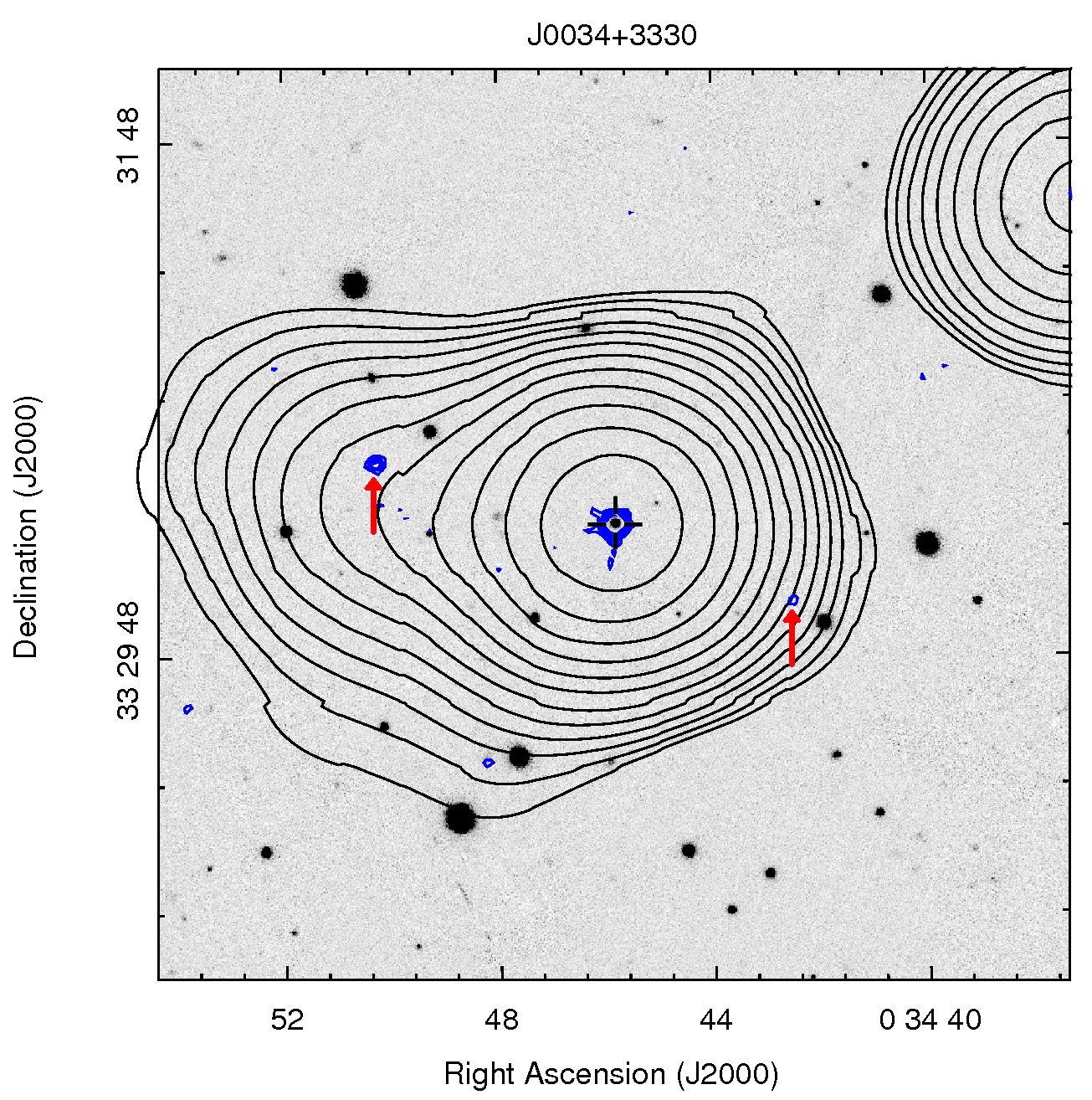}
    \includegraphics[width=0.68\columnwidth]{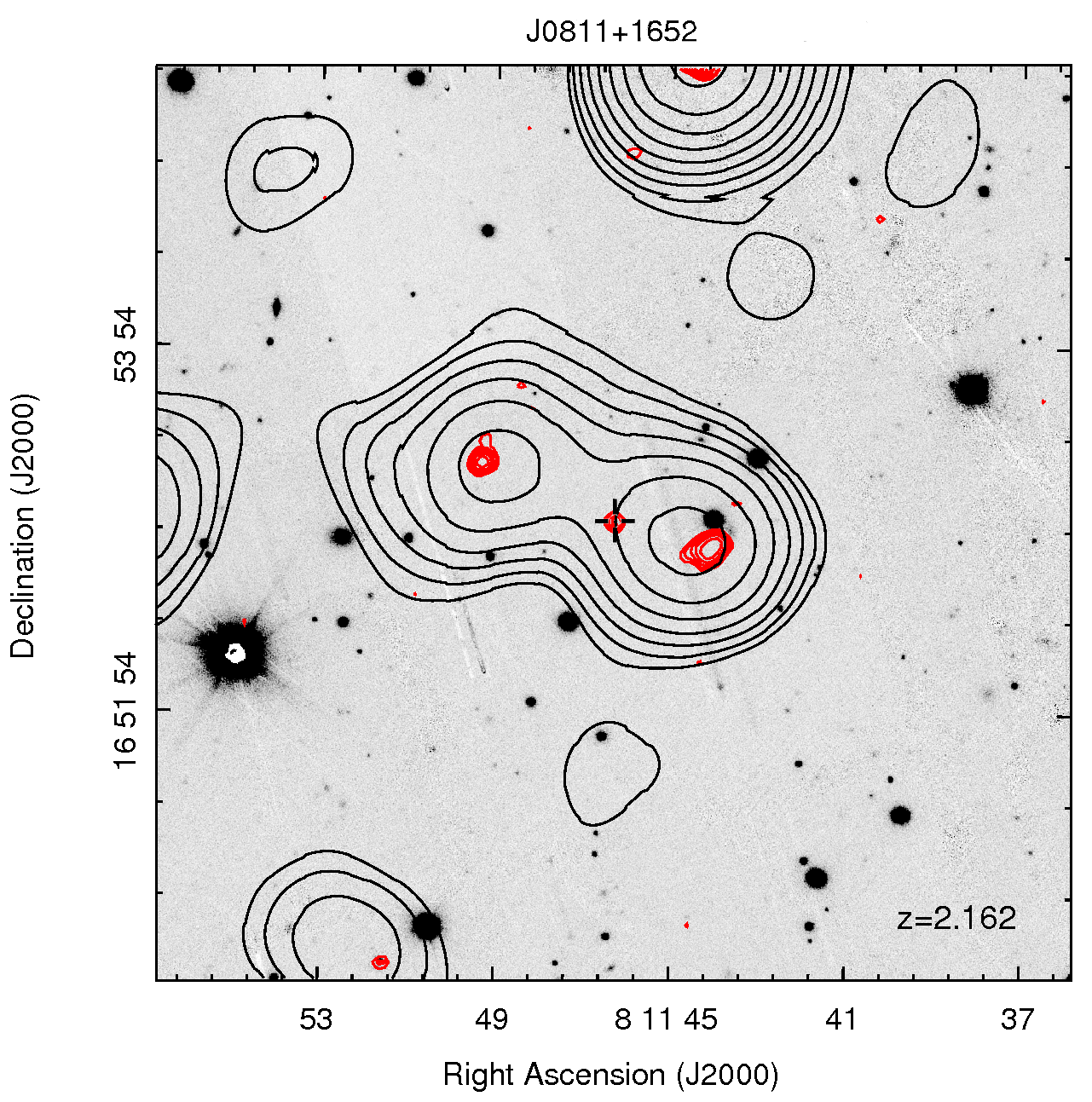}
    \includegraphics[width=0.68\columnwidth]{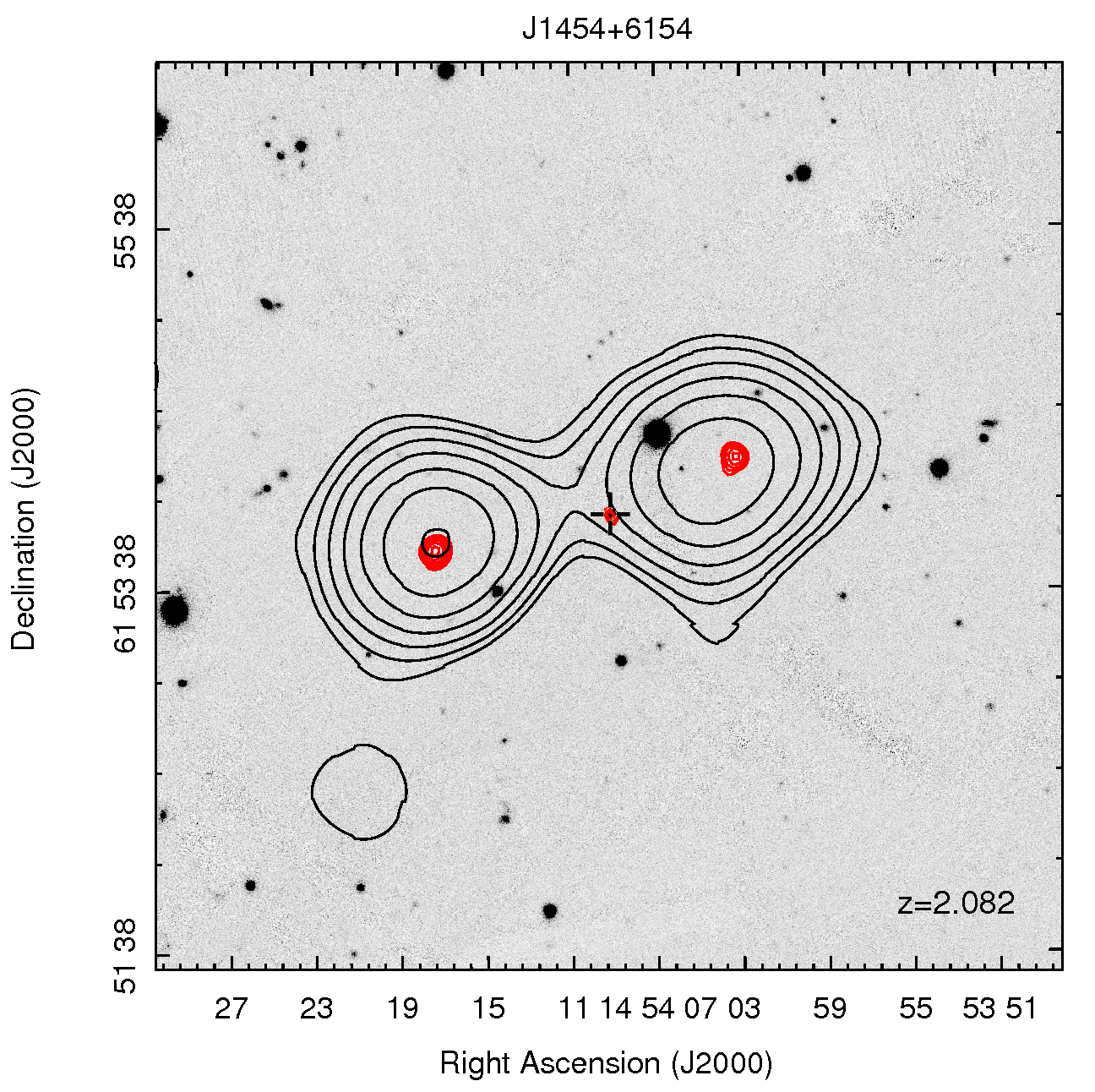}
    \includegraphics[width=0.68\columnwidth]{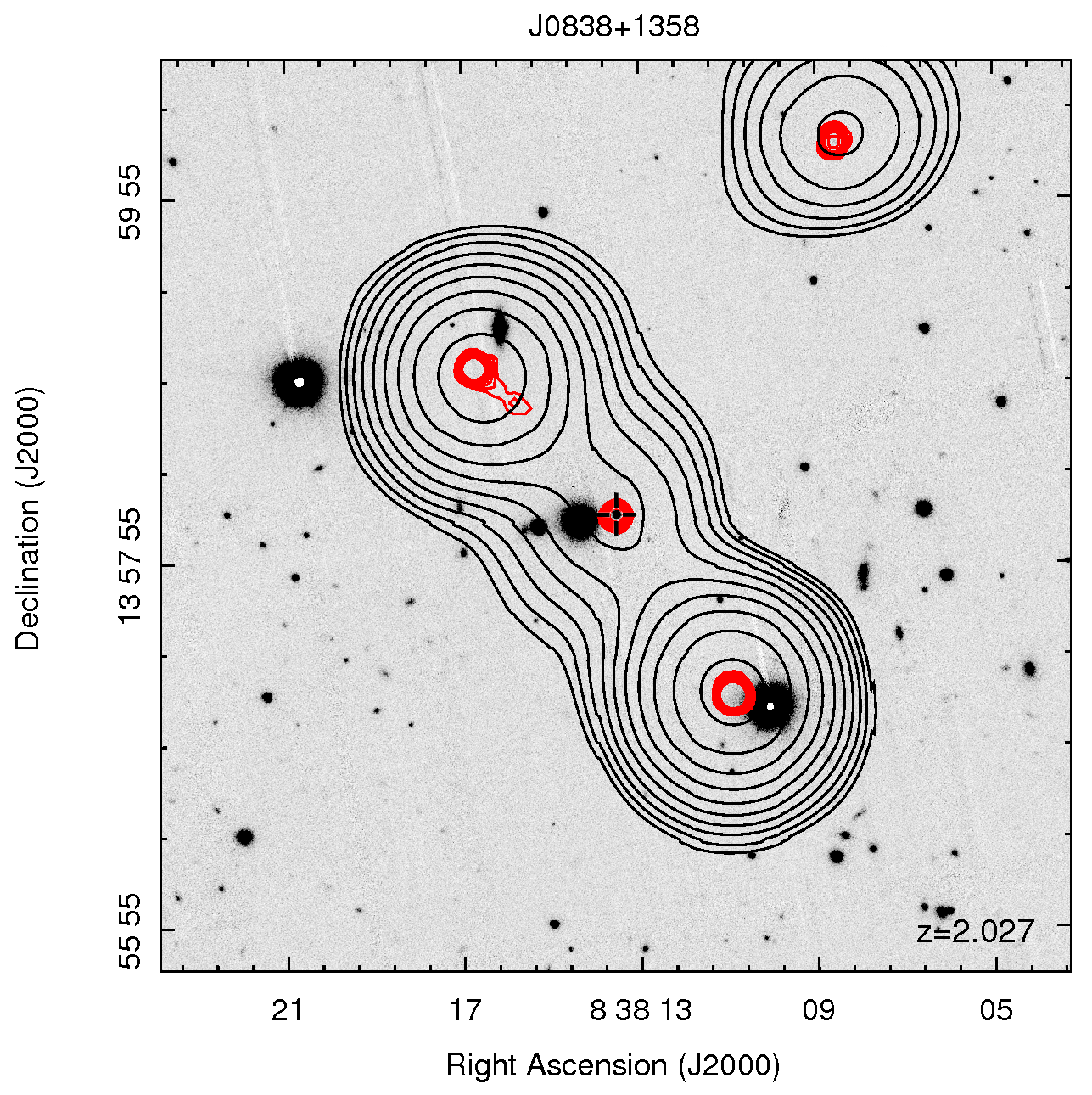}
    \includegraphics[width=0.68\columnwidth]{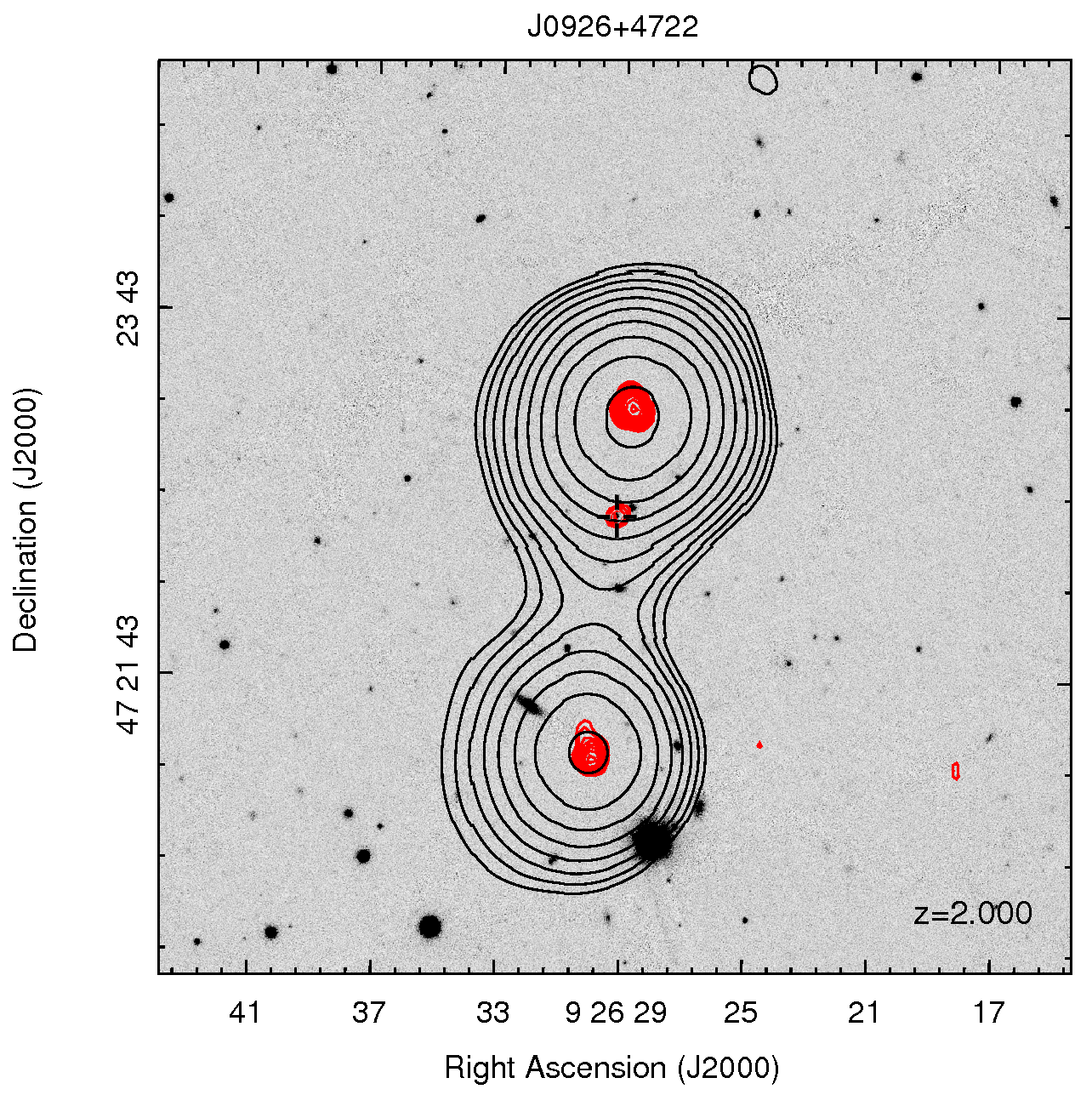}
\caption{Radio maps of new GRQs located at z$\geq$2. The NVSS black contours are overlaid onto r-band Pan-STARRS optical images. The FIRST radio contours are plotted in red and the VLASS contours in blue. The cross marks the position of the parent QSO and the red arrows in the J0034+3301 image mark the position of VLASS hot-spots}.
\label{z}
\end{figure*}

\item X-ray detections. It is expected that for more distant sources the X-ray-to-radio flux ratio increases due to higher energy density of the cosmic microwave background (CMB). At higher redshift the relativistic electrons in radio jets preferentially lose energy due to scattering on CMB photons (e.g. \citealt{simionescu2016}, \citealt{ghisellini2014}). For quasars included in our samples 72 (of the 174) GRQs and 96 (of the 367) SRQs have detections in the ROSAT all-sky survey \citep{voges2000, voges1999}, while 10 GRQs and 18 SRQs are listed in the Third XMM-Newton Serendipitous Source Catalog \citep{rosen2016}. The most distant GRQs from our sample have not been detected in X-rays to date, so they constitute good targets for future X-ray observations.\\
 
\end{itemize}

\subsection{One-sided radio jets}
\label{jet}
Some quasars from our samples show evidence of a one-sided radio jet visible very close to the host quasar. In the samples of GRQs and SRQs we found 7 and 26 such quasars, respectively. The presence of one-sided radio jets indicates the Doppler boosting of radio emission due to high radio source inclination of radio jets. Therefore, the projected sizes of such radio sources may be highly underestimated. The quasars with visible one-sided radio jet are marked with letter ``j'' in Tables 1 and 2.

\section{Radio core prominence parameter}

For the quasars from our samples, we determined the radio core prominence parameter (f$_{\rm c}$) defined as the ratio between the core flux density ($S_{\rm core}$) and the extended radio emission flux density ($S_{\rm ext}=S_{\rm tot}-S_{\rm core}$), $\rm f_c=S_{\rm core}/S_{\rm ext}$, (named as R parameter in \citealt{orr1982}). It was postulated by many authors that f$_{\rm c}$ is a good indicator of radio source orientation. A high value of the f$_{\rm c}$ parameter indicates that the radio source jets are oriented closer to the line of sight. The dependence of optical properties on the f$_{\rm c}$ parameter can result from relativistically beamed radiation or anisotropically emitted radiation due to obscuration in some directions or a line-emitting region which is not spherically symmetric \citep{jackson1991}, therefore it can be used to probe the structure of central regions in AGNs.

In our study, the f$_{\rm c}$ parameter was estimated only for those radio quasars for which we were able to separate the radio core from the extended radio emission. We did not use quasars with no radio core detection in FIRST catalogue. The radio core flux density was measured on FIRST radio maps, which have a better resolution than NVSS, while the total flux densities were measured on NVSS maps to avoid loosing weak and diffuse radio emission of radio lobes.  In the sample of GRQs we determined f$_{\rm c}$ for 225 GRQs, of which 18 have f$_{\rm c}$$>$1. An f$_{\rm c}$ value larger than one means that the radio core dominates the overall luminosity of the radio source. In the sample of SRQs we measured f$_{\rm c}$ for 284 quasars, of which only 6 have f$_{\rm c}$$>$1.
 
\begin{figure}
\centering
    \includegraphics[width=1\columnwidth]{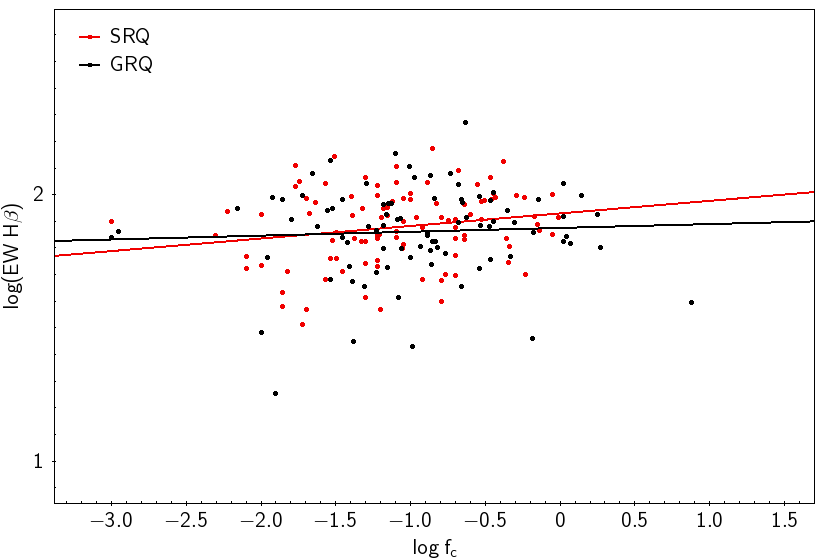}
\caption{The equivalent width of the H$\beta$ line versus the radio core prominence parameter f$_{\rm c}$. The fitted lines were obtained via the standard linear regression model. The same method was applied in Figures \ref{ewOIII}, \ref{p1} and \ref{p2}.}
\label{ew}
\end{figure}

We checked the correlations between the radio core dominance and equivalent widths (EWs) of different emission lines. The measurements of emission lines were done by \cite{suvendu2019}, where the authors estimate spectral parameters for all quasars in DR14Q. It was predicted by different models that the emission line properties depend on orientation. Such a dependence can be caused by obscuration by a clumpy torus, Doppler-boosting of continuum emission or inclination of the accretion disk (\citealt{nenkova2008}, \citealt{browne1987}, \citealt{netzer1987}). We obtained the strongest correlation between $\log \rm EW(\rm H\rm\alpha)$ and $\log \rm f_c$ (the correlation coefficient of linear fit (C) is C=0.57 for GRQs and C=0.49 for SRQs), but it should be noted that we found the H$\alpha$ line to be present in the spectra of only a few quasars, so this result is not representative for the entire sample. The EW of $\rm H\rm\beta$ broad emission line is very weakly correlated with f$_{\rm c}$ (C=0.06 for GRQs and C=0.19 for SRQs), however the overall trend of increasing EW(H$\rm\beta$) with f$_{\rm c}$ can be seen  (Figure \ref{ew}). Also other broad emission lines, like H$\rm\gamma$, MgII and CIV, are very weakly correlated with f$_{\rm c}$. In AGN models, it was predicted that the broad emission lines, which are produced close to the accretion disk, should be orientation-dependent and emitted anisotropically (e.g. \citealt{jackson1991}). The lack of correlation between broad emission lines and f$_{\rm c}$ parameter indicates that EW does not depend on orientation or that both the emission line and optical continuum show the same dependence on orientation. On the other hand, the f$_{\rm c}$ parameter may be not a good indicator of orientation for very extended radio sources because the aged radio lobes will have significantly decreased surface brightness due to radiation losses and adiabatic expansion. Also jet interactions with the intergalactic medium, as well as cosmological surface brightness dimming may affect the observed flux of extended radio structures in GRQs. It is also possible that in the case of quasars with a very high f$_{\rm c}$, the high core flux density can be caused by recurrent radio jet emission, which cannot be resolved in too low resolution radio maps (e.g. 4C +02.27, \citealt{kuzmicz2017}).

\begin{figure}
\centering
    \includegraphics[width=1\columnwidth]{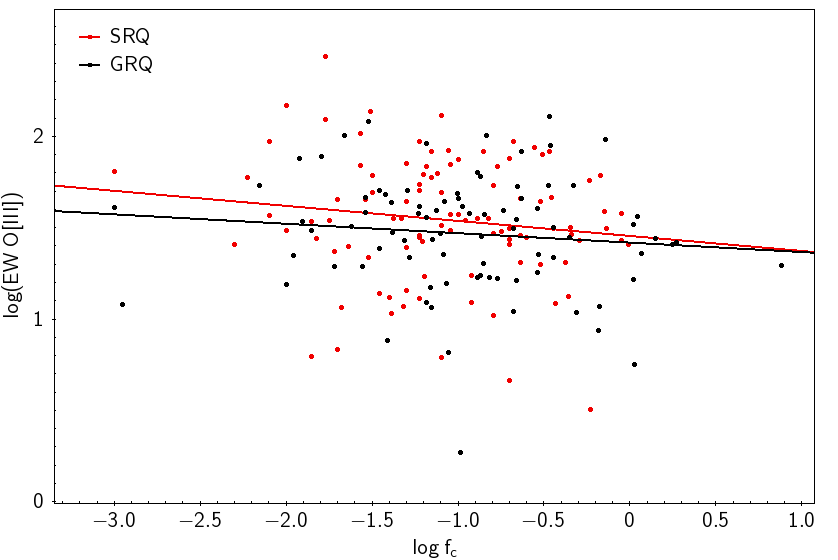}
\caption{The equivalent width of the [OIII] line against f$_{\rm c}$ parameter. }
\label{ewOIII}
\end{figure}

For our samples of quasars, we do not find either the strong anticorrelation between EW of [OIII] and f$_{\rm c}$ (Figure \ref{ewOIII}) observed by other authors (\citealt{baker1997}, \citealt{jackson1989}, \citealt{jackson2013}). The anticorrelation is predicted by the model in which the broad line region (BLR) is photo-ionized by radiation from the accretion disk and the strength of emission lines is correlated with the intensity of this radiation. When our line of sight is near perpendicular to the plane of the obscuring torus of the AGN (high f$_{\rm c}$), a low EW of [OIII] is expected. For larger inclinations (smaller f$_{\rm c}$), a higher EW should be observed due to obscuration of the ionizing component. The very weak anticorrelation for our GRQs and SRQs (C$\sim$-0.12) may have resulted from a small number of core-dominated quasars in our samples and because the f$_{\rm c}$ parameter may be not a good indicator of radio source orientation. Another possibility is that for very extended radio quasars, the structure of innermost parts of AGN (the dusty torus) does not lead to anisotropic obscuration of the ionizing continuum source. 

\section{[OIII] luminosity - radio luminosity relation}

\begin{figure}
\centering
    \includegraphics[width=1\columnwidth]{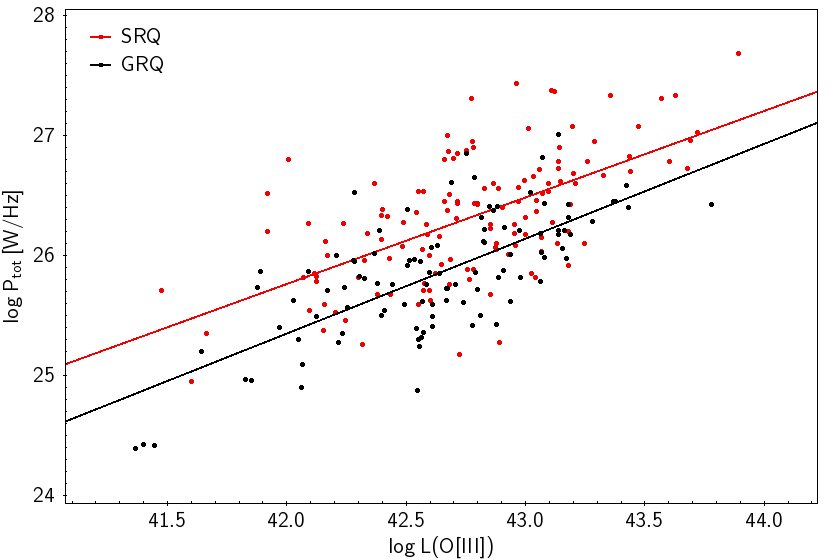}
\caption{Relation of the total radio luminosity at 1.4 GHz and the luminosity of [O III].}
\label{p1}
\end{figure}

\begin{figure}
\centering
    \includegraphics[width=1\columnwidth]{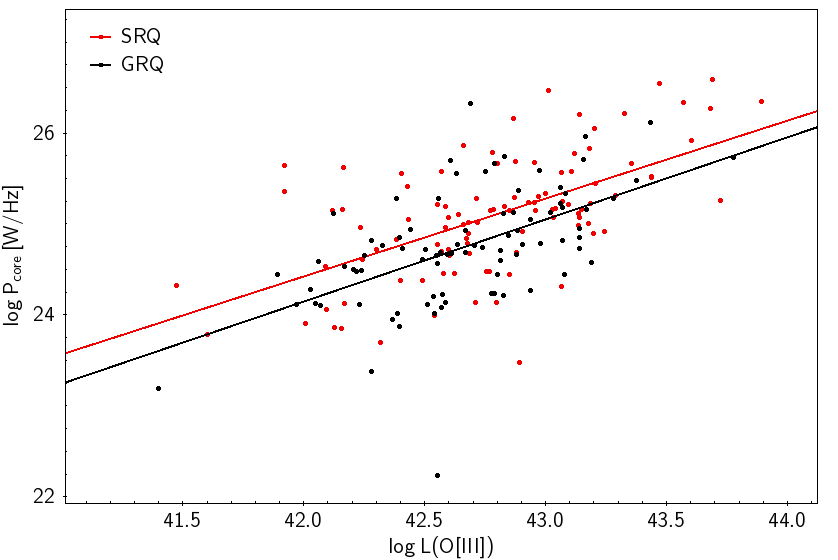}
\caption{Relation of the core radio luminosity at 1.4 GHz and the luminosity of [O III].}
\label{p2}
\end{figure}

For our samples of QSOs, we checked the relation between [OIII] luminosity (L[OIII] values are taken from \citealt{suvendu2019}) and total radio luminosity P$_{\rm tot}$, as well as the core radio luminosity P$_{\rm core}$ at 1.4 GHz (Figure \ref{p1} and \ref{p2}), which indicates a possible connection of radio jet emission with the narrow line region (NLR). For the sample of GRQs we obtained a relatively high level of correlation. For the L[OIII] vs P$_{\rm tot}$ relation, the correlation coefficient C is 0.71 and for L[OIII] vs. P$_{\rm core}$ C=0.57. A lower, but still high level of correlation, was found in the sample of SRQs. The correlation coefficients of the L[OIII] vs. P$_{\rm tot}$ and L[OIII] vs. P$_{\rm core}$ relations are C=0.61 and C=0.53 respectively. Also, when we consider the relation between L[OIII] and radio luminosity of extended radio emission (P$_{\rm ext}$=P$_{\rm tot}$-P$_{\rm core}$), there are considerable correlations in both the samples, much stronger than those obtained by \cite{gaur2019} for a sample of radio-loud quasars (C=0.25) or by \cite{tadhunter1998} for the 2 Jy complete sample of radio sources (C=0.38). It may indicate that the connection between radio emission and the NLR in GRQs is quite significant. In the standard quasar illumination model (e.g. \citealt{rawlings1991}) the correlation between radio and [OIII] luminosity is explained as due to a direct relation between the power of the photoionizing continuum and the radio jet power.

\section{Eigenvector 1}

\begin{figure}
\centering
    \includegraphics[width=1\columnwidth]{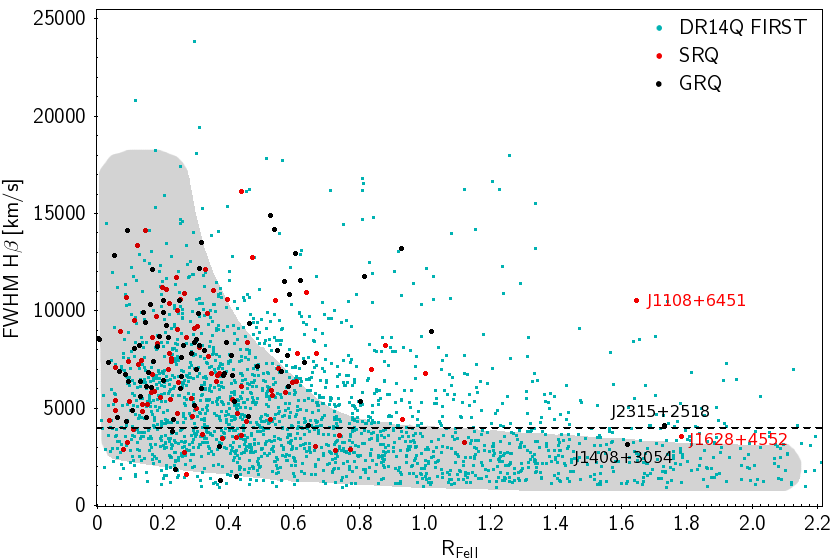}
\caption{The Eigenvector 1 plane for GRQs (black dots), SRQs (red dots) and DR14Q quasars with FIRST counterparts (DR14Q FIRST; green dots). The horizontal line at FWHM $\rm H\beta$=4000 km/s separates two different populations of quasars: population A below FWHM $\rm H\beta$=4000 km/s and population B above the FWHM $\rm H\beta$=4000 km/s. The locations of the four extreme cases J1108+6451, J1628+4552, J1408+3054, J2315+2518 are marked. The shaded area indicatively traces the distribution of a quasar sample from \cite{zamfir2010}, defining the quasar main sequence.}
\label{EW}
\end{figure}

A powerful tool for probing AGN properties and the fundamental correlations between different spectral components is Eigenvector 1 (EV1; \citealt{boroson92}). The EV1 optical plane is defined by the FWHM of the broad H$\beta$ line and the ratio R$_{\rm FeII}$ between the EW of FeII (4435--4685\AA) and the EW of the H$\beta$ broad line. The location of an object in the plane of EV1 traces the so-called quasar main sequence. In the EV1 plane two different spectral types of QSOs can be distinguished: population A with FWHM $\rm H\beta \leq$4000 km/s and population B with FWHM $\rm H\beta>$4000 km/s. Sources which belong to the same spectral type have similar spectroscopic properties, e.g. line flux ratios and line profiles \citep{sulentic2002}. The R$_{\rm FeII}$ parameter traces the variation of Eddington ratio L/L$_{\rm Edd}$, while the FWHM of H$\beta$ traces a change in orientation (\citealt{marziani2001}, \citealt{sulentic2017}). It was postulated by \cite{zamfir2008} and \cite{marziani2018} that the separation line between populations A and B at FWHM H$\beta$=4000 km/s can correspond to a critical change in accretion disk structures: from wind-dominated geometrically thick accretion disk in population A to the geometrically thin disk-dominated population B.\\
In Figure \ref{EW}, we plotted the EV1 plane for GRQs and SRQ samples. All the spectral parameters used in this study were taken from \cite{suvendu2019}. In Figure \ref{EW}, we also plotted the location of all 18 273 quasars which are flagged in DR14Q as quasars with FIRST counterparts within 2$^{\prime\prime}$ \citep{paris2018}. We refer to this sample as the DR14Q FIRST quasars. In drawing the diagram we used only the quasars that have spectra with S/N$>$5 and we do not remove GRQs and SRQs from the DR14Q FIRST sample as they constitute only 1.6\% and 2\%, respectively, of the latter sample. In Figure \ref{EW}, we also marked the division line between population A and population B at FWHM $\rm H\beta$=4000 km/s. It can be seen that most GRQs (91\%), and SRQs (83\%) belong to the B population. The objects in this region are massive, old galaxies with low accretion rates. The location of GRQs is typical for lobe-dominated quasars \citep{zamfir2008}, however few sources are located far off the main sequence area occupied by the lobe-dominated radio quasars. Two GRQs have extreme values of R$_{\rm FeII}$: J2315+2518 and J1408+3054 as are two smaller-sized radio quasars: J1628+4552 and J1108+6451. 
The location of J1408+3054 can be explained by a large amount of out-flowing gas \citep{marziani2018}, as this GRQ is a broad absorption line quasar. Spectral properties, as well as properties in other wavebands have to be studied in more detail to understand the specific positions of the remaining quasars in the EV1 plane.  \\

\begin{figure}
\centering
    \includegraphics[width=1\columnwidth]{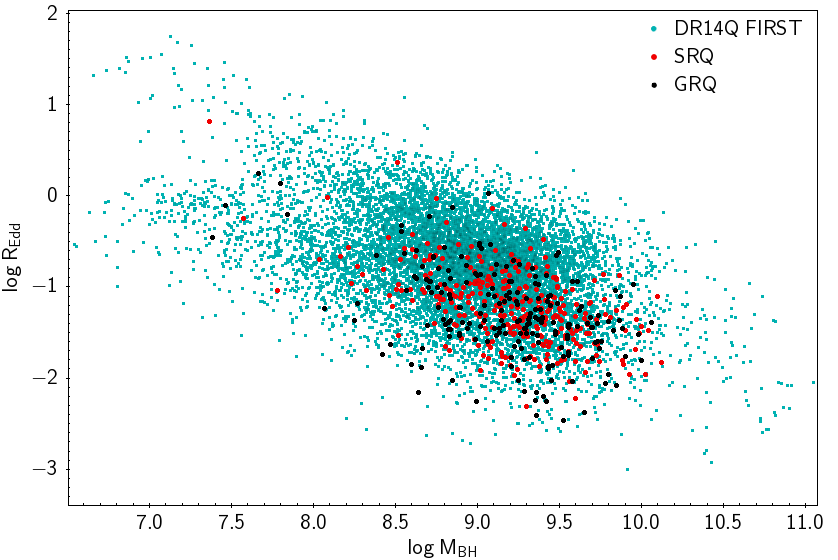}
\caption{Eddington ratio vs. BH masses for GRQs, SRQs, and DR14Q FIRST quasars. The symbols are as in Figure \ref{EW}.}
\label{bh}
\end{figure}

The location of GRQs and SRQs in the EV1 plane is in agreement with the relation between BH masses (M$_{\rm BH}$) and Eddington ratio (the ratio of bolometric to Eddington luminosity, R$_{\rm Edd}$=L$_{\rm bol}$/L$_{\rm Edd}$) plotted in Figure \ref{bh}. For all quasars studied in this paper, the estimation of BH masses, as well as the bolometric and Eddington luminosities, were taken from \cite{suvendu2019}. It can be clearly seen that GRQs and SRQs concentrate in the region of larger M$_{BH}$ and lower Eddington ratio as compared to the DR14Q FIRST quasars, which confirms that the extended radio quasars are evolved sources where accretion processes are not currently significant, representing later stages of evolution.

\section{WISE colour diagram}
In this section, we study the infrared colours based on WISE magnitudes in W1, W2, and W3 bands at 3.4, 4.6 and 12 $\mu$m respectively \citep{wright2010}. W1-W2 and W2-W3 colours were used by \cite{wright2010} for classification of astrophysical objects. Different classes of objects occupy different parts on the colour-colour diagram. In our analysis we used infrared magnitudes given in the DR14Q catalogue, which are a result of cross-matching AllWISE Source Catalog \citep{cutri2013} with DR14Q FIRST quasars. We used only the magnitudes with A or B quality flags as listed for the 198 126 AllWISE matches we found for the 526 356 DR14Q quasars. 
  
\begin{figure}
\centering
    \includegraphics[width=1\columnwidth]{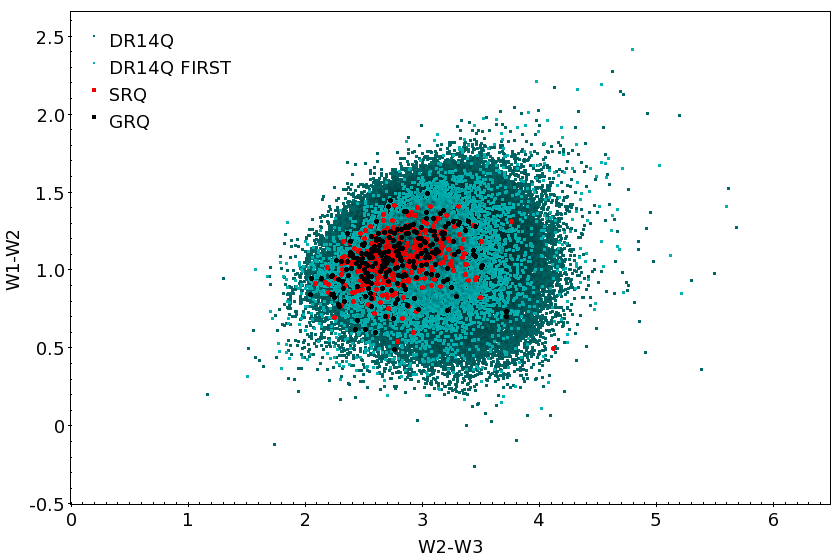}
\caption{WISE colour--colour diagram for GRQs (black dots), SRQs (red dots), DR14Q (dark green dots), DR14Q FIRST (green dots).}
\label{wise}
\end{figure}

\begin{figure}
\centering
    \includegraphics[width=1\columnwidth]{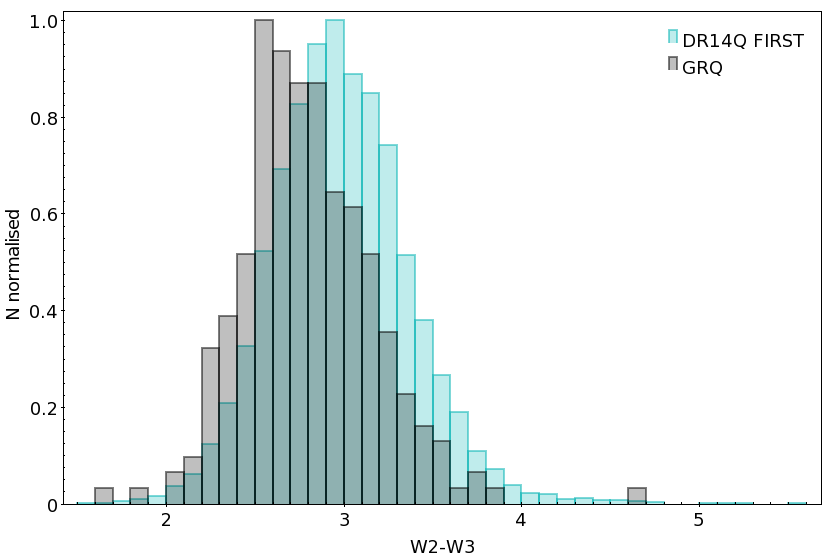}
    \includegraphics[width=1\columnwidth]{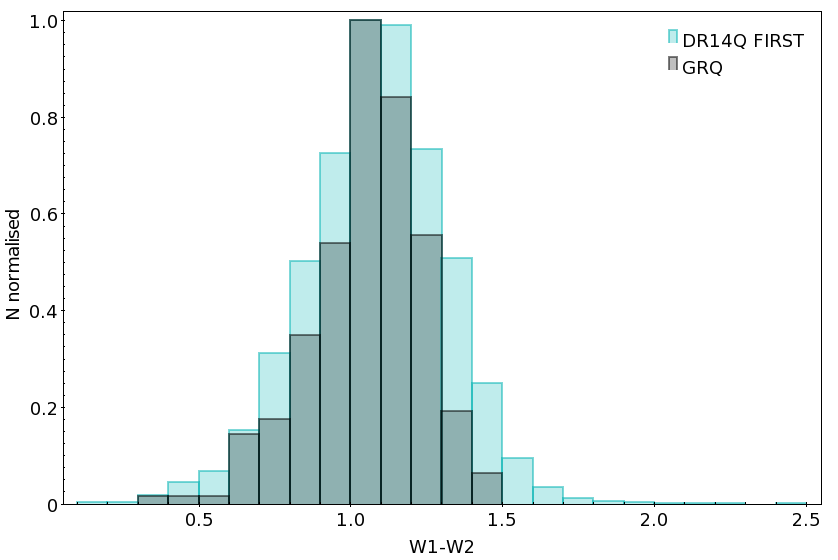}
\caption{Normalized distributions of WISE colours for GRQ and SDSS DR14Q FIRST samples.}
\label{wise2}
\end{figure}

Almost all the quasars from the GRQ and SRQ samples were detected by WISE. We compared their colours with those of the quasars from DR14Q to check their position on the colour-colour diagram relative to the entire quasar population. In Figure \ref{wise}, we can see that most quasars occupy the region centred at W1-W2$\approx$1.1 and W2-W3$\approx$3.1. Such a position within the colour-colour diagram (according to \citealt{wright2010}, \citealt{klindt2019}, \citealt{wu2012}) is characteristic for quasars. It can be seen that GRQs and SRQs occupy a smaller area than that of the quasars from DR14Q and that of the DR14Q FIRST quasars with FIRST detections, being concentrated within an ellipsoidal area in the upper left part of the diagram, with the central points of W1-W2=1, W2-W3=2.7 and the semi-major and semi-minor axes 0.8 and 0.3 respectively. The smaller area occupied by GRQs and SRQs coincides with the region of highest QSO density in DR14Q, and thus the region with the highest probability of finding quasars. However, when comparing the distributions of WISE colours for GRQs and DR14Q FIRST quasars (Figure \ref{wise2}), it is evident that GRQs have bluer W2-W3 colours, while W1-W2 colours for both the samples have similar distributions. The same behaviour is observed for the SRQ sample, which means that, as compared to the population of DR14Q FIRST quasars, GRQs and SRQs show a deficit of mid-infrared radiation towards longer wavelengths. This can indicate differences in the structure of the dusty torus (e.g. \citealt{wildy2018}) which is responsible for mid-infrared re-emission of absorbed radiation from the central source. It can also be a result of lower star formation rate in GRQs and SRQs, as compared to DR14Q FIRST quasars (e.g. \citealt{klindt2019}).
   
\section{Concluding remarks}

In this study we present the basic properties of a sample of 272 GRQs, the largest yet constructed and comprised of 98 previously known ones, plus 174 GRQs newly discovered here using NVSS, FIRST and SDSS data. Among the new GRQs, there are 70 quasars at redshift z$>$1, of which 9 are located at z$>$2, with the most distant one, J1411+0156 at z=2.946.
A large number of GRQs found by our thorough and time-consuming visual inspection of radio images proves that some of the modern/computer searches should be significantly modified. We enlarged the number of known GRQs nearly threefold, showing that previous catalogues of GRSs are incomplete and future extensive search is also necessitated for GRGs.\\
The search method used for the identification of GRQs also allowed us to compile a sample of 367 extended radio quasars with smaller radio structures (0.2$<$D$<$0.7 Mpc), which can be used in other investigations. It contains only the quasars found in this work.\\

Our analysis of optical, radio and infrared host properties of GRQs, SRQs and quasars from the DR14Q FIRST sample gave the following results:
\begin{itemize}
\item The radio core prominence for GRQs and SRQs is just very weakly correlated with the EW of broad emission lines. We did not find either any statistically significant anticorrelation between EW of [OIII] and radio core prominence.    
\item We obtained a strong correlation between the [OIII] luminosity and the total and core radio luminosities for GRQs, as well as SRQs.
\item Based on the position of quasars within the Eigenvector 1 plane, most of GRQs and SRQs belong to B population with FWHM $\rm H\beta$$>$4000 km/s, representing evolved objects with high BH masses and the low accretion rates.  
\item There are few quasars in our samples with extreme values of EW FeII to EW H$\beta$ ratio, compared to other GRQs and SRQs. They should be studied separately to understand their specific location in the Eigenvector 1 plane. 
\item The positions of GRQs and SRQs on the WISE colour-colour diagram show a deficit of mid-infrared radiation towards longer wavelengths, which may indicate differences in the structure of the dusty torus, as compared to quasars from DR14Q FIRST sample.  
\end{itemize} 
We found no significant differences between GRQs and SRQs, which show that in general the extended radio quasars are a group of objects with similar spectral and infrared properties. The size of radio structures seems to be independent of the host galaxies' spectral properties, while there is a strong connection between the radio luminosity and the [OIII] emission line emitted in the NLR. Also the infrared W2-W3 colours of both GRQs and SRQs have smaller values than those of DR14Q quasars in general, indicating differences in their dusty torus structure. 

\section{Acknowledgments}

We thank the referee for valuable comments and corrections that helped to improve the paper. We also thank Conor Wildy for his helpful comments.\\
This paper was supported by the National Science Centre, Poland through the grant 2018/29/B/ST9/01793.

\clearpage

\startlongtable
\begin{deluxetable}{c c c c c c c c c c}
\tablecaption{List of 174 newly discovered GRQs.}
\tablecolumns{10}
\tablenum{1}
\tablecomments{
{\bf Column description:} (1) -- source name, (2) and (3) -- J2000.0 source coordinates, (4) -- redshift, (5) -- angular size; measured on FIRST maps as a distance between hot-spots, (6) -- projected linear size, (7) -- radio core prominence, (8) core radio luminosity at 1.4 GHz in units of W$\slash$Hz, (9) -- total radio luminosity at 1.4 GHz in units of W$\slash$Hz, (10) -- flag: (fn) -- angular size for radio sources with FIRST radio structure visible only on the one side; measured between the FIRST hot-spot on the one side and the maximum of NVSS emission on the other side, (nn) -- angular size for radio sources with no visible FIRST hot-spots; measured between maxima of NVSS emission, (v) -- angular size for radio source with no visible FIRST hot-spots; measured between hot-spots detected in VLASS, (s) -- the total radio luminosity is affected by a background radio source. We subtracted 8 mJy VLASS 3 GHz contaminating source assuming the spectral index $\alpha$=$-$0.7, (c) -- candidate GRQ with uncertain radio morphology, (h) -- uncertain host position, the galaxy SDSS J142302.67$+$265014.9 located at redshift z$_{\rm phot}$=0.7 is also a possible host, (j) -- the quasars with visible one-sided radio jet (Section \ref{jet}). \\
}
\tablehead{
\colhead{IAU}	&\colhead{$\alpha$(2000)}&\colhead{$\delta$(2000)} &\colhead{z}	& \colhead{LAS$_{\rm FIRST}$}	& \colhead{D$_{\rm FIRST}$}	&\colhead{$\rm f_c$}	&\colhead{$\log$P$_{\rm core}$}	&\colhead{$\log$P$_{\rm tot}$} & \colhead{Flag}\\
\colhead{name}	&\colhead{(h m s)}	&\colhead{($\rm^{o}$ $^{\prime}$ $^{\prime\prime}$)}&\colhead{}	&\colhead{(arcmin)}       &\colhead{(Mpc)}	& \colhead{}    &\colhead{W Hz$^{-1}$}	&\colhead{W Hz$^{-1}$}	&\\
\colhead{(1)}	&\colhead{(2)}		&\colhead{ (3)}					& \colhead{(4)}	&\colhead{(5)}            & \colhead{(6) }      & \colhead{(7)} & \colhead{(8)}       &\colhead{(9)}	&\colhead{(10)}
}
\startdata
J0012$+$0902	&	00 12 09.85	&	$+$09 02 49.3	&	0.7930	&	1.90	&	0.85	&	0.06	&	24.85	&	26.18	&   \\
J0023$-$0926	&	00 23 34.07	&	$-$09 26 24.3	&	0.7886	&	2.71	&	1.22	&	0.03	&	24.14	&	25.76	&   \\
J0024$+$0057	&	00 24 22.62	&	$+$00 57 51.2	&	0.9646	&	1.50	&	0.72	&	0.02	&	23.84	&	25.74	&   \\
J0032$+$2327	&	00 32 10.51	&	$+$23 27 55.0	&	0.6393	&	2.83	&	1.17 	&	-	&	-	&	25.49	& v \\
J0032$+$1852	&	00 32 17.09	&	$+$18 52 56.1	&	0.5307	&	2.99	&	1.13 	&	-	&	-	&	25.73	& v \\
J0032$-$0605	&	00 32 46.62	&	$-$06 05 33.9	&	0.7540	&	3.96	&	1.75 	&	0.34	&	25.28	&	25.95	& fn \\
J0034$+$3330	&	00 34 45.88	&	$+$33 30 18.6	&	2.2270	&	1.71	&	0.86 	&	-	&	-	&	27.29	& v \\
J0043$+$3137	&	00 43 59.87	&	$+$31 37 20.3	&	0.6266	&	3.01	&	1.23 	&	-	&	-	&	26.24	& v \\
\enddata
\end{deluxetable}
\begin{figure}
\end{figure}
\clearpage

\newpage                                                                           
\startlongtable
\begin{deluxetable*}{c c c c c c c c c c}
\tablecaption{List of 367 newly found SRQs.}
\tablecolumns{10}
\tablenum{2}
\tablecomments{The column description and symbols meaning as in Table 1.\\
Angular size of sources which can not be measured in FIRST and VLASS radio maps are measured to the 3$\sigma$ contour level in NVSS maps, therefore their sizes may be overestimated. Such quasars are marked by (*).
}
\tablehead{
\colhead{IAU}	&\colhead{$\alpha$(2000)}	&\colhead{$\delta$(2000)} &\colhead{z}	& \colhead{LAS$_{\rm FIRST}$}	& \colhead{D$_{\rm FIRST}$}	&\colhead{$\rm f_c$}	&\colhead{$\log$P$_{\rm core}$}	&\colhead{$\log$P$_{\rm tot}$} & \colhead{Flag}  \\
\colhead{name}	&\colhead{(h m s)}	&\colhead{($\rm^{o}$ $^{\prime}$ $^{\prime\prime}$)}&\colhead{}	&\colhead{(arcmin)}       &\colhead{(Mpc)}	& \colhead{}    &\colhead{W Hz$^{-1}$}	&\colhead{W Hz$^{-1}$} & \\
\colhead{(1)}	&\colhead{(2)}		&\colhead{ (3)}					& \colhead{(4)}	&\colhead{(5)}            & \colhead{(6) }      & \colhead{(7)} & \colhead{(8)}       &\colhead{(9)} & \colhead{(10)}
}
\startdata
J0001$+$2452	&	00 01 26.18	&	$+$24 52 09.5	&	1.4230	&	1.34	&	0.68	&	-	&	-	&	26.36	& v  \\
J0007$-$0051	&	00 07 17.82	&	$-$00 51 37.8	&	0.5259	&	1.66	&	0.62	&	0.03	&	23.70	&	25.26	& fn \\
J0010$+$3329	&	00 10 25.63	&	$+$33 29 36.8	&	0.7420	&	1.23	&	0.54	&	-	&	-	&	27.00	& nn \\
J0014$+$0039	&	00 14 28.74	&	$+$00 39 04.7	&	1.1630	&	1.00	&	0.50	&	0.12	&	25.27	&	26.34	&   \\
J0015$+$0400	&	00 15 53.74	&	$+$04 00 39.6	&	0.7590	&	1.23	&	0.55	&	0.10	&	25.12	&	26.23	&   \\
J0022$+$2833	&	00 22 16.36	&	$+$28 33 54.9	&	0.5846	&	1.55	&	0.66	&	0.04	&	25.16	&	25.59	&   \\
J0022$-$0145	&	00 22 44.30	&	$-$01 45 50.9	&	0.6899	&	1.55	&	0.66	&	0.04	&	25.16	&	26.62	&   \\
J0029$+$1527	&	00 29 28.20	&	$+$15 27 22.4	&	0.7094	&	1.55	&	0.67	&	-	&	-	&	25.53	&  nn \\
\enddata
\end{deluxetable*}
\footnotesize
\scriptsize  
\begin{figure}
\end{figure}


\end{document}